\documentclass[12pt]{article} % Change to "article"

\usepackage[utf8]{inputenc}   % Encoding
\usepackage{amsmath, amssymb} % Math support
\usepackage{graphicx}         % Include graphics
\usepackage{siunitx}          % SI units
\usepackage{geometry}         % Page geometry
\usepackage{cleveref}         % Intelligent cross-referencing
\usepackage{authblk}

\crefname{figure}{Fig.}{Figs.}
\crefformat{figure}{#2Fig.~#1#3}  % No space between "Fig." and the number
\usepackage{float}            % Enhanced floating objects
\usepackage{caption}          % Customizing captions
\usepackage{booktabs}         % Better table formatting
\usepackage[backend=biber,style=numeric,citestyle=numeric,sorting=none]{biblatex} % Reihenfolge der Zitate im Text
\usepackage{todonotes}

\newcommand{\GaAs}{\ensuremath{GaAs}}
\newcommand{\AlAs}{\ensuremath{AlAs}}
\newcommand{\AlGaAs}{\ensuremath{Al_{0.15}Ga_{0.85}As}}
\newcommand{\nm}{\nano\metre}
\newcommand{\um}{\micro \metre}
\newcommand{\mm}{\milli \metre}

\DeclareUnicodeCharacter{0308}{ö}
\title{Bright quantum dot light sources using monolithic microlenses on gold back-reflectors}

\author[1]{Moritz Langer}

\author[1]{Sai A. Dhurjati}

\author[1]{Yared G. Zena}

\author[1]{Ahmad Rahimi}

\author[1]{Mandira Pal}

\author[1]{Liesa Raith}

\author[1]{Sandra Nestler}

\author[2]{Riccardo Bassoli}

\author[3]{Frank H. P. Fitzek}

\author[4]{Oliver G. Schmidt}

\author[1,2]{Caspar Hopfmann}
\affil[1]{Institute for Emerging Electronic Technologies, IFW Dresden, Helmholtzstraße 20, 01069 Dresden, Germany}
\affil[2]{Quantum Communication Networks research group, Deutsche Telekom Chair of Communication Networks, Dresden University of Technology, Germany}
\affil[3]{Deutsche Telekom Chair of Communication Networks, Dresden University of Technology, Germany}
\affil[4]{Research Center for Materials, Architectures and Integration of Nanomembranes (MAIN), Chemnitz University of Technology, Chemnitz, Germany}

\begin{document}

\date{\today}  % Leave empty to avoid displaying the date

\maketitle

\begin{abstract}
We present the fabrication process of bright $\GaAs$ quantum dot (QD) photon sources by non-deterministic embedding into broadband monolithic $\AlGaAs$ microlens arrays on gold-coated substrates. Arrays of cylindrical photoresist templates, with diameters ranging from \qty{2}{\um} to \qty{5}{\um}, are thermally reflowed and subsequently transferred into the $\AlGaAs$ thin-film semiconductor heterostructure with embedded quantum dots through an optimized anisotropic and three-dimensional shape-preserving reactive ion etching process. This methodology facilitated the fabrication of large-scale (\qty{2}{\mm} $\times$ \qty{4}{\mm}) and densely packed arrays of uniformly shaped microlenses ($\sim$ \qty{40e3}{\per \square\mm}), with the brightest emissions from QDs embedded in microlenses exhibiting lateral diameters and heights of \qty{2.7}{\um} and \qty{1.35}{\um}, respectively.
Finite-difference time-domain simulations of both idealized and fabricated lens shapes provide a comprehensive three-dimensional analysis of the device performance and optimization potentials such as anti-reflection coatings. It is found that free-space extraction (fiber-coupled) efficiencies of up to \qty{62}{\percent} (\qty{37}{\percent}) are achievable for hemispherical QD-microlenses on gold-coated substrates. A statistical model for the fabrication yield of QD-microlenses is developed and experimentally corroborated by photoluminescence spectroscopy of fabricated microlens arrays. This analysis exhibited a free-space intensity enhancement by factors of up to $\times$ \num{200} in approximately \num{1} out of \num{200} microlenses, showing good agreement to the theoretical expectations.
This scalable fabrication strategy underscores the potential of these compact, high-efficiency sources offering new prospects for applications of these devices in future large-scale quantum networks.
\end{abstract}

% \keywords{Microlens, Quantum dot, Bright Entangled Photon Pair Source, Compact photon source}

\section{\label{introduction} Introduction}

Efficient quantum emitters capable of producing single photons and entangled photon pairs on demand are essential for the advancement of quantum communication networks \cite{Al-Amri2016-ga, lo2014secure, reiserer2015cavity, borregaard2019quantum} and cryptographic systems \cite{gisin2002quantum}. While conventional sources of entangled photon pairs often rely on spontaneous parametric down-conversion, quantum emitters, particularly $\GaAs$ quantum dots (QDs), offer superior potential performance due to their deterministic behavior \cite{chen2018highly, Wang2019, Liu2019, hopfmann2021maximally} and integrability with semiconductor technologies. Among various quantum emitters, QDs have demonstrated exceptional promise as sources for fast and deterministic entangled photons in practical quantum network applications \cite{heindel2012quantum, schimpf2021quantum}.

In order to facilitate the wide-spread adaption of quantum light sources in future large-scale quantum communication networks, these sources need to exhibit competitive performance across a set of key performance indicators such as clock rate $\Gamma_{\text{rep}}$, entanglement fidelity $f$, photon indistinguishability $I$, single-photon purity $g^{(0)}$, and source efficiency $\eta$ \cite{Loock2020}. While some of these parameters have been successfully addressed in individual experiments, integrating all of them into a single source remains a significant challenge \cite{hopfmann2021maximally}. A major limitation in current condensed matter sources is the extraction efficiency—the ratio of photons collected to those generated—which is typically low due to the omnidirectional emission characteristics of solid-state quantum emitters. In semiconductor-based sources this issue is exacerbated by the significant refractive index mismatch between the emitter’s host material and the surrounding medium (air), causing photon extraction to be severely limited by total internal reflection. Consequently, efficiencies of bulk semiconductor quantum light sources are typically below $\qty{2}{\percent}$ without additional photonic engineering \cite{zwiller2002improved, Shields2007}. Photonic micro-structuring offers a promising solution to improve extraction efficiency, either by enhancing coupling to specific emission modes through the Purcell effect in micro cavities or by reducing the internal reflection at the semiconductor-air interface through optical microscale engineering.

Examples for the employment of efficient photon sources based on micro cavities are micropillar \cite{ding2016demand, somaschi2016near, moczala2019strain, gines2022high}, fiber  \cite{najer2019gated}, and photonic crystals \cite{kim2016two, kors2017telecom} cavities. It has been demonstrated, that these micro resonator structures with large quality (Q) factors are very suitable for high brightness single photon sources \cite{Tomm2021}. Due to the narrow-bandwidth of these cavities, they are however not ideal for multi-photon emission, such as required for entangled pair sources based on single semiconductor QDs \cite{Shields2007}. In recent years, circular Bragg cavities, which feature moderate Q-factors and very small mode volumes, have been employed to achieve Purcell-enhanced emission of entangled photon pairs using emission from quantum dot exciton-biexciton cascades \cite{Wang2019, Liu2019}. The main drawback of the circular Bragg cavities is however that they require very precise positioning - down to a few nanometers - of the micro cavity structure with respect to the QD in order to preserve the high entanglement fidelities \cite{xu2022bright}.  

In terms of geometrical approaches for efficient light extraction, solid immersion lenses (SILs), fabricated from materials like epoxy \cite{trojak2018combined}, dielectrics \cite{bishop2022enhanced}, or III-V compounds \cite{chen2018highly, nie2021experimental}, are one prominent example. These lenses offer broadband enhancement and can be fabricated through simple methods like stick-and-glue assembly or 3D printing \cite{sartison2017combining}, yielding photon extraction improvements up to two orders of magnitude. However, challenges related to scalability and fabrication precision remain. For example, the small membrane areas limit scalability, while the fabrication process requires careful adjustment of the membrane-SIL gap and polishing of high-refractive index material SILs \cite{chen2018highly, yang2020quantum, nie2021experimental}.

Recent advancements have focused on integrating microlenses monolithically into the emitter's host material, eliminating the need for precise gap adjustments and issues with thermal expansion mismatch between the materials \cite{sardi2020scalable}. The small sizes of these monolithic microlenses offers significant advantages for photon extraction in terms of device scalability, on-chip integration as well as direct coupling to fiber optics \cite{liu2024broadband, li2021wet, schmidt2019deterministically, Schlehahn2018, Musial2020,Rickert2025,Langer2025}. Initial devices were designed using in-situ grown distributed Bragg reflectors (DBRs) as rear mirrors, though these mirrors offer moderate performance due to their limited acceptance angles of about $\SI{19}{\degree}$ and narrow stop bands \cite{gschrey2015highly}. Metallic mirrors, on the other hand, offer superior broadband reflectivity, reduced device sizes and do not require the complex layer optimization associated with DBRs \cite{liu2024broadband}.

\section{\label{sec:concept} Concept}

\begin{figure}[H]
    \centering
    \includegraphics[width=\textwidth]{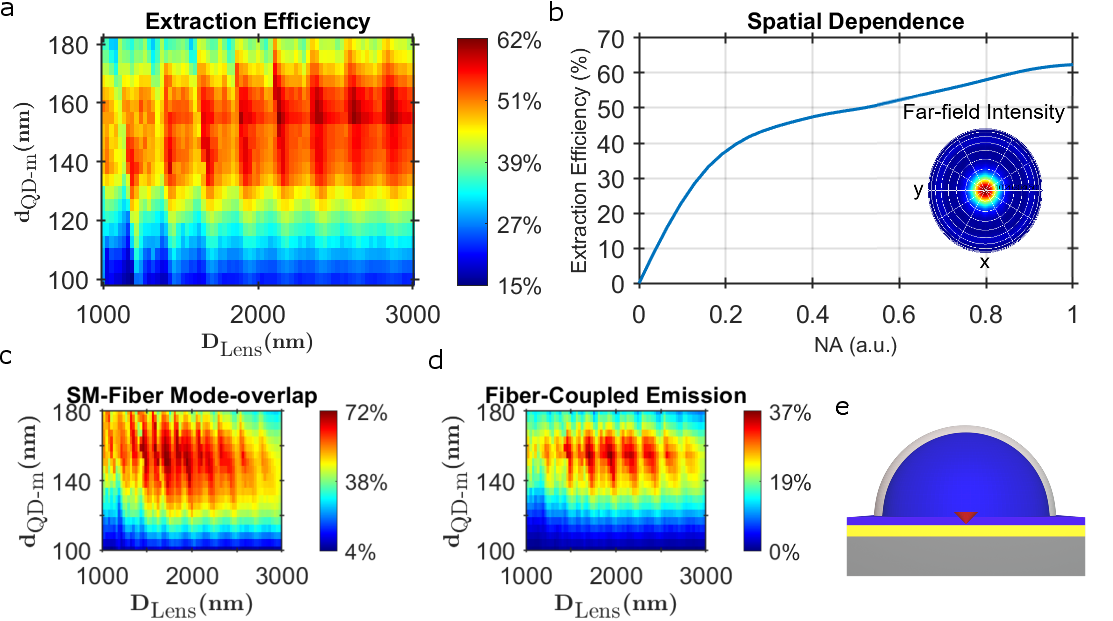}
    \caption{Finite-difference time-domain (FDTD) simulation results of a hemispherical microlens with anti-reflection coating on a gold-coated substrate. 
    a) Extraction efficiency at a numerical aperature $\text{NA}=1$ for a QD at the microlens center, analyzed by varying the QD to gold mirror distance $d_{QD-m}$ and microlens diameter $D_{lens}$. 
    b) Extraction efficiency vs NA for a QD at $d_{QD-m}=\qty{155}{\nm}$ and $D_{\text{lens}}=\qty{2135}{\nm}$. The angular far-field distribution is shown as an inset. 
    c) Mode-overlap of QD-microlens emission with respect to the spatial distribution of a lensed 780HP single-mode fiber at wavelength \qty{780}{\nm} and $\text{NA}= 0.6$. 
    d) Fiber-coupled QD emission into single-mode fiber given by the product of a) and c). 
    e) Schematic illustration of the simulation model: $\GaAs$ substrate (gray), \qty{500}{\nm} gold back-reflector (yellow), $\AlGaAs$ membrane with thickness $d_{QD-m}$ of \qty{100}{\nm} to \qty{180}{\nm} and hemispherical $\AlGaAs$ microlens diameter $D_{lens}$ of \qty{1}{\um} to \qty{3}{\um} (blue), \qty{112}{\nm} $\mathrm{Al_{2}O_{3}}$ anti-reflection coating (light-gray), and $\GaAs$ QD source (red) emitting at \qty{780}{\nm}.}
    \label{fig:concept}
\end{figure}

Given the constraints and considerations of high-performance semiconductor QD based entangled photon pair sources outlined above, in this work, we explore a scalable microlens fabrication approach of monolithic microlenses on Van-der-Waals bonded QD-nanomembranes to gold-coated $\GaAs$ substrates. This approach allows the creation of free-standing microlenses, optimized for photon extraction of centrally embedded QDs, while simplifying the fabrication process to facilitate high yields and performance characteristics \cite{ma2015highly}. Due to the lack of strong resonant effects, the monolithic microlenses feature broadband operation, low strain and reduced positioning requirements while enabling scalable high-density microlens device arrays. Through simulations of $\GaAs$ QDs integrated into $\AlGaAs$ microlenses, we investigate the impact of emitter positioning on photon extraction and infer obtainable fabrication yields. Additionally, we highlight the importance of anti-reflection coatings \cite{schnauber2015bright} in further improving extraction efficiency and mitigating fabrication precision constraints. These innovations pave the way for scalable, quantum emitters based on monolithic QD-microlenses suitable for broadband industrial adaption.

In order to benchmark the maximal performance of the concept of monolithic microlens based QD photon sources, it is useful to consider an idealized microlens design. In order to minimize the losses of back-reflection on the semiconductor-air interface, the angle of the light-ray incident should be perpendicular. Due to the small size of the quantum dot with respect to its emission wavelength, cf. \cref{sec:methods}, it can in good approximation be modeled as an omnidirectionally emitting point source. From this consideration, it follows that the natural shape from maximal light extraction is that of a spherical lens with the QD at its center. Since the sample is only accessible from the top hemispherical direction, the lower hemisphere can be replaced by a reflecting (i.e. metallic, such as gold) layer. In order for the $\GaAs$ QD to efficiently confine carrier states, it has to be embedded in a $\AlGaAs$ matrix, cf. \cite{keil2017solid}. Therefore, a minimal distance of $>$ \qty{50}{\nm} between the reflecting layer interface and the quantum dot is required. In order to reduce the residual reflection at the $\AlGaAs$-air interface, an antireflection coating (ATC) of materials such as $Al_2O_3$ can be employed. The resulting idealized monolithic microlens device design is illustrated in \cref{fig:concept}e.

To evaluate the extraction efficiency of the chosen device design, a comprehensive investigation of hemispherical microlenses is conducted using finite-difference time-domain (FDTD) simulations. By using these investigations the extraction efficiency (EE), the lensed single-mode fiber to emission mode overlap (CE) and the QD emission into a lensed single-mode fiber (ICE), defined as the product of EE and CE, are evaluated analogous to Ref. \cite{nie2021experimental}. For the evaluation of the overlap of the QD emission with a mode from a single-mode fiber (ICE), a 780HP fiber lensed by an objective with a numerical aperture (NA) of \num{0.6} is assumed. The results are shown in \cref{fig:concept}a-d as a function of the microlens diameter $D_{lens}$ and distance between gold-back-reflector and QD $d_{QD-m}$. The $\mathrm{Al_{2}O_{3}}$ anti-reflection layer features a thickness of \qty{112}{\nm}. The QD is modeled as a centrally positioned point dipole source, emitting at a wavelength of \qty{780}{\nm}, see also \cref{fig:concept}e. The device parameters of  $d_{QD-m}$ and $D_{lens}$ are optimized for maximal ICE rather than EE, the reason is that the QD-microlens photon sources are to be combined with direct in-situ coupling to single mode fibers, as performed in Ref. \cite{Langer2025}. Further details regarding the performed FDTD simulations can be found in \cref{sec:methods}. 

For microlenses within the depicted range $D_{lens} = \qty{1}{}-\qty{3}{\um}$, the total spatial EE, shown in \cref{fig:concept}a, demonstrates a periodic modulation with a period of approximately \qty{250}{\nm}, aligning with the Fabry–P\'erot resonances of the microlens. On the $d_{QD-m}$ axis, a single maximum corresponding the QD position co-location to a Fabry–P\'erot resonance maxima can be observed at $\qty{155}{\nm}$. The maximal value of the EE at an NA of \num{1} is about \qty{62}{\percent} at values of $D_{lens}$ close to \qty{2.9}{\um}. In \cref{fig:concept}c the determined mode overlap CE, exhibits similar behavior to EE but peaks at a value of \qty{72}{\percent} at $D_{lens} = $ \qty{1.7}{\um}. Consequently, the total emission into a single mode fiber ICE, shown in \cref{fig:concept}d, is maximal at a value of \qty{37}{\percent} for $D_{lens}$ and $d_{QD-m}$ values of \qty{2.0}{\um} and \qty{155}{\nm}, respectively. The angular extraction efficiency of the optimal ICE device is depicted in \cref{fig:concept}b. By investigating extended ranges of parameters of both $D_{lens}$ and $d_{QD-m}$, not shown here, it has been verified that the presented optimal ICE device represents a global maximum for the chosen idealized device design.
In summary, the simulations of the idealized hemispherical QD-microlens device design reveal optimum efficiencies in both light extraction and coupling, characterized by symmetric and concise far-field emission properties, thus indicating the potential to achieve emission efficiencies as high as \qty{37}{\percent} into single-mode fibers. Any deviation from the ideal hemispherical lens shape will decrease this optimal value as well as alter the specific $D_{lens}$ and $d_{QD-m}$ values.

\section{\label{sec:fabrication} Fabrication}

In order to realize bright QD-microlenses close to the idealized hemispherical design outlined above, a series of optical lithography steps are performed on a layered $GaAs$/$AlAs$ heterostructure embedded with droplet etched $\GaAs$ QDs. The semiconductor heterostructure wafers are obtained by epitaxial growth using a molecular beam epitaxy (MBE) machine, cf. \cite{keil2017solid}. Details of the heterostructure design and fabrication can be found in \cref{sec:methods}.
The fabrication method for monolithic QD-microlens arrays consists of five sequential steps, cf. \cref{fig:reflow}a-f. Initially, a protective resist layer is applied using spin-coating and subsequently dried. Following this, the $\AlGaAs$ membrane with embedded $\GaAs$ quantum dots is released by selective chemical etching of the $\AlAs$ sacrificial layer using hydrofluoric (HF) acid. Consequently, the membrane is transferred onto a gold-coated $\GaAs$ substrate, which serves as a highly reflective rear mirror for the microlenses. Thereafter, the protective resist layer is removed, and cylindrical resist structures are deposited in a new resist layer through photolithography. These structures are subjected to reflow on a hot-plate, yielding photoresist microlens templates. Subsequently, the photoresist templates are etched into the membrane using anisotropic reactive-ion etching (RIE). This technique permits substantial adaptability of the resulting monolithic microlens shape through the modulation of resist templates and etching process. The reflow and shape transfer processes are discussed in the following, further details on the membrane transfer process are given in \cref{sec:methods}.

\subsection{\label{sec:reflow}Reflow}
\begin{figure}[H]
    \centering
    \includegraphics[width=\textwidth]{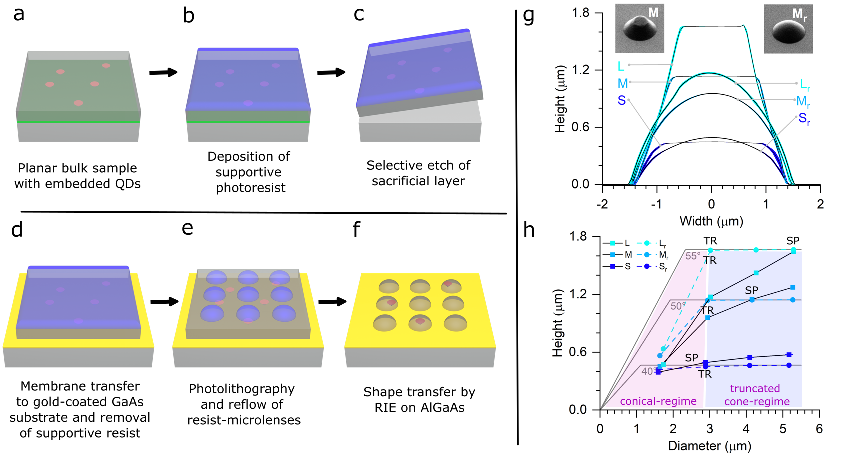}
    \caption{Fabrication process of monolithic microlenses by thermal reflow of photoresist templates with diameters of \qty{2}{\um}–\qty{5}{\um} on gold-coated $\GaAs$ substrate.
    a) QD sample with $\GaAs$ bulk (gray), sacrificial layer (bright green), and $\AlGaAs$ active region (dark-gray transparent) with embedded liquid-droplet etched QDs (red). 
    b) Deposition of AZ1500 photoresist layer (blue) to support the membrane in the transfer process. 
    c) Selective etching of $\AlAs$ sacrificial layer to detach $\AlGaAs$ membrane from $\GaAs$ substrate. 
    d) Van-der-Waals bonding of $\AlGaAs$ membrane to gold-coated $\GaAs$ substrate. 
    e) Resist removal and second deposition of AZ1500 photoresist for lithography of micro-mesa array and reflow into lens shaped resist templates (blue). 
    f) Photoresist pattern transfer onto a monolithic $\AlGaAs$ membrane via reactive-ion etching, removing $\AlGaAs$ and QDs from unprotected areas, and leaving monolithic microlenses with embedded QDs.
    g) Averaged AFM height profiles of 15 photoresist structures with \qty{3}{\um}-diameter and varying spin-coated resist thicknesses ($L$:\qty{1.66}{\um}, $M$:\qty{1.14}{\um}, $S$:\qty{0.45}{\um}), measured pre- and post-reflow ($r$), its standard deviation is indicated by blue shading. SEM images show structure M before ($M$) and after reflow ($M_{r}$).
    h) Average thickness and diameter of resist structures pre- (dotted colored lines) and post-reflow (solid black lines). Gray lines represent ratios for ideal conical resist structures with bevel angles of \qty{40}{\degree}, \qty{50}{\degree}, and \qty{55}{\degree}. The red shaded area represents the regime with photoresist in a conical shape and blue shaded area for truncated cone shape, with the transition-diameter TR and the settle-points SP, signifying when the heights of reflow resist microlenses becomes larger than the resist thickness.}
    \label{fig:reflow}
\end{figure}

The simulations discussed in \cref{sec:concept} identify an optimal ICE performance for $D_{lens}$ in the range of \qty{1.7}{\um} and \qty{2.7}{\um}. To achieve these diameters practically, resist templates with initial diameters between \qty{2}{\um} and \qty{5}{\um} are sufficient, taking into consideration the anisotropic effects and diameter reduction due to isotropic etching. Exact adjustments of the resist height-to-radius aspect ratio are essential for the realization of hemispherically shaped structures. Accordingly, resist thicknesses between \qty{0.4}{\um} to \qty{2}{\um} are investigated, under the presumption of $\GaAs$ to photoresist RIE etching selectivity ratios in the range of \num{2} and \num{5}.

To investigate the fabrication and thermal reflow process, cylindrical resist templates with diameters of \qty{2}{\um}, \qty{3}{\um}, \qty{4}{\um}, and \qty{5}{\um}, and resist thicknesses of $S$: \qty{0.45}{\um}, $M$: \qty{1.14}{\um}, and $L$: \qty{1.66}{\um} are used. The averaged thickness profiles of ten \qty{3}{\um} resist templates before and after thermal reflow measured by atomic force microscopy (AFM), cf. \cref{sec:methods}, are depicted in \Cref{fig:reflow}h. Temperature treatments exceeding the resist-specific softening temperature (for AZ1500 ca. \qty{130}{\celsius}) induce reflow, thereby transforming the structures into lens-like shapes due to the inherent tendency of the liquid resist minimizing its free surface energy \cite{ONeill2002}. A comparison of the profiles before and after reflow is presented in \cref{fig:reflow}g, corroborated by SEM images shown as insets. Further pre- and post-profiles used to derive the curves of \cref{fig:reflow}h are shown in supplemental \cref{fig:resistfabrication_all}. The analysis reveals, contrary to the nominal cylindrical resist structures with vertical sidewalls, that pre-reflow templates exhibit beveled sidewalls and resemble truncated cones. This morphology is attributed to photolithographic limitations in resolution, i.e. the proximity effect.\\

For resist templates with diameters exceeding \qty{2}{\um}, the pre-reflow structures resemble truncated cones, whereas smaller diameters undergo a transition to true conical shapes. Within the conical-shape regime, the post-reflow height becomes largely independent of the initial resist thickness, being predominantly influenced by the template diameter. For resist thicknesses of $>$ \qty{0.45}{\um}, this transition is observed within the range of \qty{2}{\um} to \qty{3}{\um}, as shown in \cref{fig:reflow}h by the declining thicknesses with reducing diameters. For thinner resist layers, the diameter limit of the truncated cone transition is reduced.

The aspect ratio of the resist templates within the conical-shape regime can be anticipated by an idealized cone model, where the contact angles of resist droplet and gold surface fall within a range akin to those observed experimentally, this is visualized by the solid gray curves of \cref{fig:reflow}h. The experimental data reveal a gradient comparable to that predicted by the cone model, although a persistent reduction in thickness is noted. This reduction is attributed to the volume loss during reflow due to evaporation, which is between \qty{10}{} to \qty{30}{\percent}. The transition diameter thereby determines the maximum attainable aspect ratio for microlenses produced in this process. Enhancing the sidewall angle through advanced, high-resolution lithography techniques could further augment the available aspect ratio range and alleviate this limitation.\\

It is noteworthy that the reflow process does not result in an increase in resist droplet diameter for $S$ and $M$, with marginal increase for $L$, suggesting that even at elevated reflow temperatures the droplet surface tension of the resist droplet is sufficient to hold its shape. This simplifies the precise control of the lateral droplet dimensions, as it is the same as that of the deposited photoresist templates. The height of the resist droplets can be controlled by varying the employed resist thickness of the photolithography process using the diameter-to-height curves shown in \cref{fig:reflow}h. For the investigated resist diameters and thicknesses, microlens resist templates of diameters in the range of \qty{1.5}{\micro \metre} and \qty{5.0}{\micro \metre} and heights between \qty{0.5}{\micro \meter} and \qty{1.7}{\micro \meter} are achieved. The obtainable aspect ratios ($\frac{Diameter}{2 \, Height}$) of the photoresist microlens templates are therefore in the range of \num{0.9} to \num{3.7}.
Notably, resist structures with aspect ratios close to 1 tend to develop wrinkles during reflow, thereby compromising their structural integrity. Meanwhile, thin resist layers offer good control of the aspect ratio with superior reproducibility due to reduced thickness variation across diameters. These properties render thin resist layers ($S$) the preferred approach to facilitate consistent and reliable microlens resist templates.

\subsection{\label{sec:shape_transfer}Shape Transfer}
\begin{figure}[ht]
    \centering
    \includegraphics[width=\textwidth]{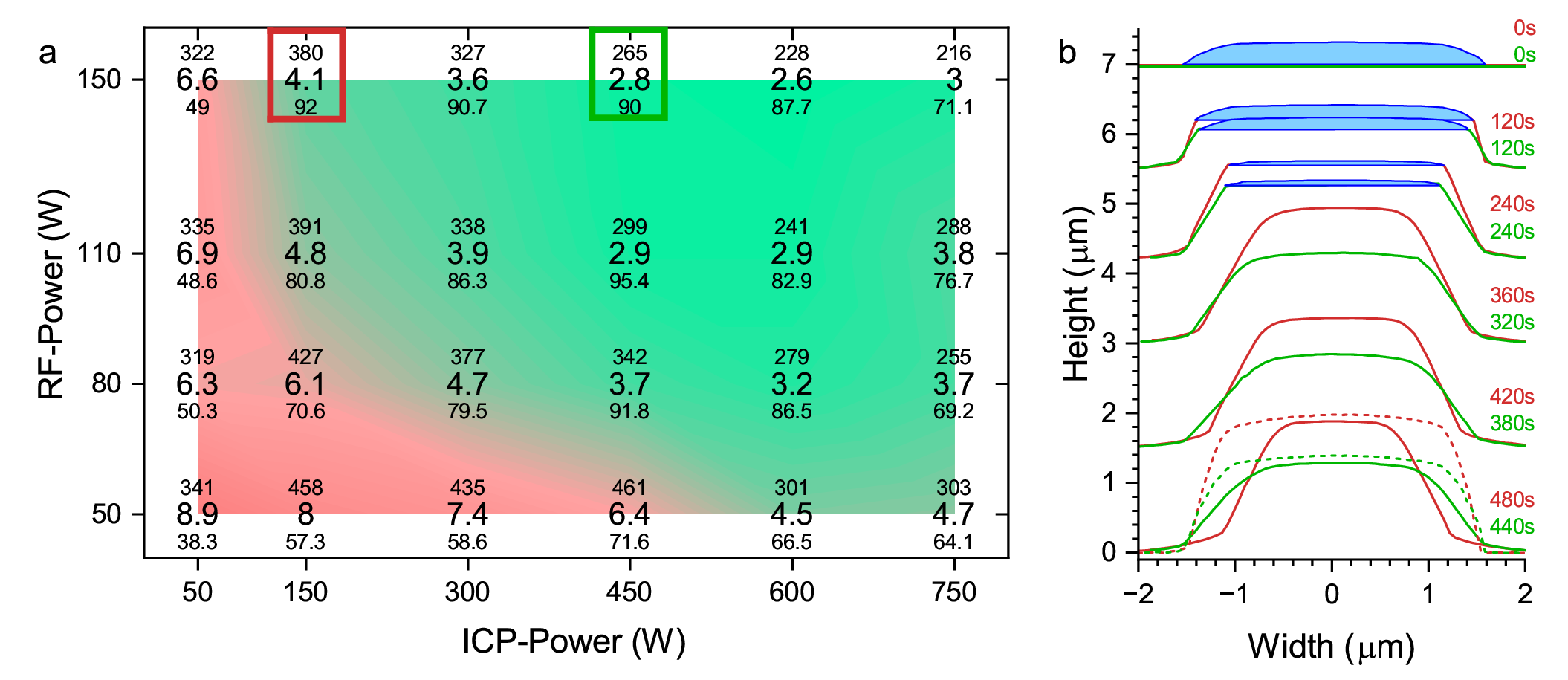}
    \caption{Characterization of shape transfer from photoresist to $\GaAs$ in an ICP-RIE process: 
    a) Average etch rates (\qty{}{\nm\per\minute}) for $\GaAs$ (top) and AZ1500 photoresist (bottom) vary with RF power (\qty{50}{\watt}–\qty{150}{\watt}) and ICP power (\qty{50}{\watt}–\qty{750}{\watt}) in a $\mathrm{Cl_{2}}$:Ar plasma. Etch selectivity (middle values, resist to $\GaAs$ rate ratio) is shown as a red-green gradient: red for high (\qty{8.9}), green for low (\qty{2.6}) selectivity. 
    b) Time-resolved etch profiles compare RF=\qty{150}{\watt} and ICP=\qty{150}{\watt} (red) with RF=\qty{150}{\watt} and ICP=\qty{450}{\watt} (green), matching red and green squares in a). The resist layers (blue) are completely removed after \qty{240}{\second}, consequently only etching $\GaAs$ substrate. Red and green dotted lines show expected etch profiles without isotropic etching.}
    \label{fig:rficpvariation}
\end{figure}

The microlens shape transfer from the resist templates into the $\AlGaAs$ membrane is achieved through a dry RIE process in conjunction with a $\mathrm{Cl/Ar}$ gas mixture. The selection of these gases is advantageous due to the volatility of the reaction products, namely $\mathrm{GaCl_{x}}$, $\mathrm{AlCl_{x}}$, and $\mathrm{AsCl_{x}}$, which inhibits the development of non-volatile compounds or passivation layers that might arise when utilizing alternative gases, such as $\mathrm{BCl_{3}}$, $\mathrm{SiCl_{4}}$, or $\mathrm{CH_{4}/H_{2}}$ mixtures \cite{juang1994comparing, volatier2010extremely}. Argon gas, despite its inertness to $\AlGaAs$, causes a sputtering effect that provides controlled and smooth surface ablation \cite{benevides2019ar}. Further details on the employed RIE etch system can be found in \cref{sec:methods}.\\

The fabrication process of monolithic microlenses markedly diverges from conventional planar etching techniques, which primarily emphasize vertical etching without considering the ablation of the resist material. In contrast, microlens fabrication necessitates a comprehensive, three-dimensional analysis of both semiconductor material and resist ablation during the etching process. A critical performance parameter in this context is the etch selectivity, defined as the ratio of the etch rate of $\GaAs$ to that of the resist. The etch rates are influenced by the composition of the plasma, which is adjusted in this study via the inductively coupled plasma (ICP) power and the accelerating radio-frequency (RF) power. The correlation between ICP and RF powers as well as the etch rates is illustrated in \cref{fig:rficpvariation}a. It should be noted that the etching rates for $\AlGaAs$ are anticipated to differ from those for $\GaAs$, depending on the aluminum concentration in the material. Nonetheless, it is expected that the qualitative behavior of the selectivity will remain consistent across varying $Al$ concentrations.\\

The data indicates a distinct trend in which elevated ICP and RF powers are associated with a decline in etch selectivity, as evidenced by the color shift from red to green, representing a transition from high to low selectivity. This reduction in selectivity is ascribed to the increase in ICP power, which increases the kinetic energy of chloride radicals and reactive ions. Consequently, this reduces the surface absorption rate and the chemical etching rate of $\GaAs$. Moreover, increased RF power further diminishes selectivity, which can be attributed to the reduced chloride absorption at the $\GaAs$ surface, as well as the enhanced physical sputtering and ablation of the resist.

In order to gain a deeper understanding of the shape transfer process, time-resolved etching experiments were conducted, cf. \cref{fig:rficpvariation}b. In these experiments, ten samples possessing identical resist structures with high aspect ratios were subjected to etching over durations ranging from \qty{0}{} to \qty{480}{\second}. The compiled height profiles reveal two distinct etching series: one at RF = \qty{150}{\watt}/ICP = \qty{150}{\watt} (red) and another at RF = \qty{150}{\watt}/ICP = \qty{450}{\watt} (green). Although both series exhibit comparable resist etch rates of $\qty{90(2)}{\nm\per\minute}$, they differ in their $\GaAs$ etch rates and etch selectivity. The complete removal of the resist occurs between $\qty{240}{\second}$ and $\qty{320}{\second}$, respectively, with overetching observed beyond this point. The final observed heights of $\qty{1890}{\nm}$ (green) and $\qty{1290}{\nm}$ (red) indicate etching selectivity ratios of about 4.1 and 2.8, respectively, thus demonstrating excellent reproducibility in agreement with \cref{fig:rficpvariation}a.\\

The etching selectivity is a crucial parameter affecting the contact angle and shape of the resulting monolithic microlens. As selectivity increases, the vertical component of shape transfer is enhanced, resulting in steeper sidewalls for a given initial resist profile. This effect enables the processing of even non-ideal hemispherical resist templates into $\AlGaAs$ microlenses that closely approximate a hemispherical shape. Under conditions of increasing ICP power, the etching process demonstrates a marked anisotropic character, as indicated by the almost unchanged diameter of the green height profiles even during overetching periods of up to \qty{120}{\second}. Conversely, lower ICP powers promote isotropic etching, leading to increased etching in both horizontal and vertical directions. The isotropy factor, defined as the ratio of horizontal to vertical etch rates, was found to be approximately \num{0.26} for RF = \qty{150}{\watt}/ICP = \qty{150}{\watt} and roughly 0.01 for RF = \qty{150}{\watt}/ICP = \qty{450}{\watt} in $\GaAs$. This variation in isotropy is responsible for the observed reduction in diameter and alteration in the profiles of microlenses subjected to longer etch times at small ICP powers. Furthermore, the isotropic etching behavior extends beyond $\GaAs$ to the resist material itself. The calculated isotropy factor for the resist was ca. \num{0.28} at RF = \qty{150}{\watt}/ICP = \qty{150}{\watt} and \num{0.18} at RF = \qty{150}{\watt}/ICP = \qty{450}{\watt}. This observation further highlights the complexity of the etching process and underscores the importance of optimized power parameters to ensure accurate transfer of the photoresist templates into the desired microlens shapes.\\

In summary, the RIE process allows for precise modulation of the shape transfer characteristics of resist microlenses. High values of resist to semiconductor etch selectivity ratios enhance the formation of steep sidewalls and can result in etched geometries that are close to hemispherical, if the initial resist microlenses feature matching aspect ratios. However, due to the diameter reduction caused by isotropic etching, the meticulous regulation of the etching duration is required for achieving precise monolithic microlens diameters. Additionally, it is crucial to minimize the likelihood of surface damage to $\GaAs$ and $\AlGaAs$ substrates, which arise from exposure to energetic argon and chloride ions during the RIE process, particularly under elevated RF power settings \cite{pang1983effects,glembocki1991optical}. This potential damage can be alleviated by utilizing low RF powers.\\

\section{\label{sec:results} Results}
\begin{figure}[H]
    \centering
    \includegraphics[width=\textwidth]{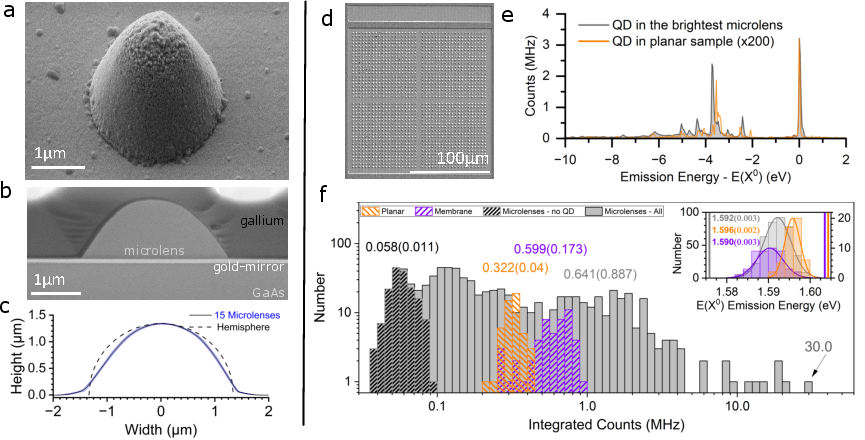}
    \caption{Shape and spectrographic analysis of fabricated microlenses. 
    a) Scanning electron beam micrograph (SEM) of a single microlens at \qty{45}{\degree} inclination. 
    b) Cross-sectional SEM image of a monolithic microlens using ion beam milling, demonstrating the seamless integration of lens and gold-coated $\GaAs$ substrate. Gallium infilling was used to stabilize the shape during ion-beam milling. 
    c) Average AFM height profiles of \num{15} microlenses (blue line), its standard deviation is indicated by blue shading. A hemispherical profile with \qty{2.7}{\um} diameter (black dotted line) is depicted for comparison. 
    d) Top-view SEM image of a \num{40} $\times$ \num{40} microlens array demonstrating scalable fabrication capability. 
    e) Low-temperature PL spectra of QD emissions at exciton saturation using a \qty{660}{\nm} continuous laser of the brightest microlens device (gray spectra) and a typical QD of a planar sample (orange spectra, 200× magnified). 
    f) Histogram of the integrated PL intensities between \qty{775}{\nm} - \qty{785}{\nm} of \num{50} QDs in planar and \num{50} QDs in membrane samples, as well as 800 microlenses from an \num{20} $\times$ \num{40} array. The statistical averages of the observed sample cohorts are denominated. The respective QD exciton ground-state ($X^0$) emission energies and their normal distribution models are shown as an inset. }
    \label{fig:luminescence}
\end{figure}

By precise control of resist profile and etching parameters, we fabricate large arrays of monolithic $\AlGaAs$ microlenses, each incorporating embedded $\GaAs$ QDs, as shown in \cref{fig:luminescence}d. The array consists of $40 \times 40$ lenses on a \( \qty{2}{}\times \qty{2}{\milli\meter\squared} \) membrane, which has been transferred from the epitaxial wafer onto a gold mirror (\cref{sec:growth}). The membrane has a thickness of \( \qty{1540}{\nano\meter} \), with the QDs positioned \( \qty{155}{\nano\meter} \) above the gold reflector. Photoresist array structures are defined by a diameter of \( \qty{3}{\micro\meter} \) and a thickness of \( \qty{440}{\nano\meter} \) and subsequently reflowed at \( \qty{140}{\degreeCelsius} \) for \( \qty{60}{\minute} \). The structures are etched with RIE at an RF power of \( \qty{110}{\watt} \) and an ICP power of \( \qty{450}{\watt} \) for \( \qty{6}{\minute} \). Following the etching process, the sample is immediately coated with a \( \qty{10}{\nano\meter} \) layer of \( \mathrm{Al_{2}O_{3}} \) to prevent it from oxidation. The spatial profiles of the fabricated microlenses are presented in \cref{fig:luminescence}a-c. A focused ion beam cross-sectional image reveals the seamless integration of the microlens with the gold layer, facilitating high precision in $d_{QD-m}$ via precise MBE growth as well as the employed membrane lift-off and placement procedure. The gold layer remains virtually unaffected by the RIE etching process and therefore functions as a reference plane for the AFM height profiles shown in \cref{fig:luminescence}c. 
AFM analysis of \num{15} randomly selected microlenses from an array indicates highly uniform lens profiles, exhibiting an absolute (relative) standard deviation in width of \(\qty{80}{\nm}\) (\(\qty{3}{\percent}\)) and height \(\qty{11}{\nm}\) (\(\qty{0.8}{\percent}\)), underscoring the precision of the fabrication process. The measured profile closely resembles that of a hemisphere at its zenith, with slight linear deviations near the meridians.\\ 

The assessment of light extraction enhancement is conducted by low-temperature high-resolution photoluminescence (PL) spectroscopy on QDs within a planar sample, a membrane sample, and \num{800} microlenses of one $20 \times 40$ field. Each QD and microlens was analyzed for its ground-state exciton ($X^0$) emission energy and integrated PL intensity, across the spectral range of \(\SI{775}{\nm}-\SI{785}{\nm}\), at excitation saturation using \qty{660}{\nm} above-band excitation. This methodology is adapted from Refs. \cite{nie2021experimental, Hopfmann2021a}. The measurement of QDs in the membrane structure and QDs in microlenses are performed on the same sample, together with a planar unprocessed heterostructure sample from the same wafer, for comparative purposes. For the characterization of the $X^0$ emission energy, only PL spectra where the exciton was the predominant emission feature were chosen. These spectra were subsequently evaluated through an automated curve modeling program and numerically integrated to determine the $X^0$ energies and QD emission intensities, respectively, cf \cref{fig:luminescence}f. 
No significant shifts in exciton emission energy were observed following sample processing. Minor deviations, as observed in the planar or membrane samples, can likely be ascribed to inherent variations in MBE-grown wafer. The epitaxial lift-off process appears to induce neither strain nor consequential energy-shifting stress. This aspect is very advantageous with respect to achieving small exciton fine structure splittings, which is beneficial for high-quality entangled photon pair sources \cite{hopfmann2021maximally}.

Asserting the extraction efficiency value of microstructures experimentally is an inherently complex and error-prone process as it requires the employment of indirect methods such as accurate simulations and determination of all systematic optical losses or demanding investigations using resonant excitation schemes \cite{hopfmann2021maximally}. For this reason, the light enhancement is investigated using a comparison of the ratio of integrated emission intensities of single monolithic QD-microlenses to single unprocessed QD samples. This ratio, called light extraction enhancement LEE, is directly accessible by photoluminescence (PL) spectroscopy, under the condition of consistency in the measurement procedure. Details on the employed PL-spectroscopy setup can be found in \cref{sec:methods}.

When comparing the LEE of a membrane to that of planar reference sample, an enhancement of $\times2$ is observed. This can be attributed to the effect of the gold mirror in reflecting approximately half of the emitted light back towards the collection side. Out of the \num{800} microlenses examined, \num{164} exhibited no QD emission, \num{359} showed integrated intensities below the average for the planar sample, and \num{277} displayed enhanced intensities. Due to the statistically distributed relative positioning of microlenses and QDs, the observed variation of QD-microlens performance is fully expected. A detailed discussion of the device yield is found below. The most luminous microlens within the array achieved an enhancement factor of $\times113$, whereas the most luminous device on this sample attained a LEE factor of $\times200$. A comparison of QD emission spectra of this particular device to that of a typical planar sample QD is provided in \cref{fig:luminescence}e. In the spectrographic investigations, it was found that the charge state of the majority of QD remained neutral, indicating negligible effects from surface damage in the etching process. This is another inherent advantage of the presented monolithic microlens design as the distance of centrally embedded QDs from the etched surface of ca. \qty{1}{\um} is relatively far compared to, for example, circular Bragg cavities.\\

\begin{figure}[H]
    \centering
    \includegraphics[width=\textwidth]{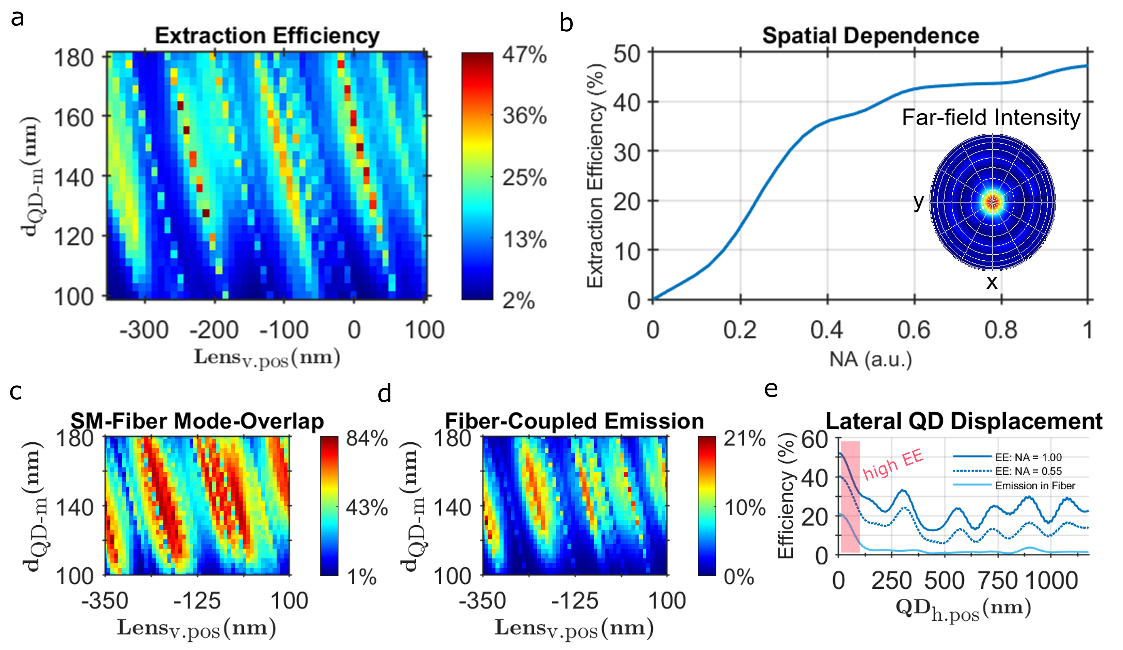}
    \caption{Finite-difference time-domain (FDTD) simulations of $\AlGaAs$ monolithic microlenses shape profiles derived from atomic force microscopy, cf. \cref{fig:luminescence}c. Device modeling and simulations are performed analogous to \cref{sec:concept}.
    a) Extraction efficiencies assuming a numerical aperture (NA) of \num{1} as a function of QD to gold mirror distance $d_{QD-m}$ and the vertical displacement of the microlens shape $Lens_{v.pos}$. 
    b) NA-dependent extraction efficiency for $d_{QD-m} = $\qty{127}{\nm} and $Lens_{v.pos} = $\qty{-212}{\nm}. The angular far-field intensity distribution of the emission is shown as an inset. 
    c) Mode overlap of QD-microlens emission profile and lensed single-mode fiber, assuming a lens NA of $0.6$.
    d) Fiber-coupled efficiency of emission from monolithic QD-microlenses into a lensed single-mode fiber as a function of $d_{QD-m}$ and $Lens_{v.pos}$. 
    e) Efficiencies for horizontal displacement of the QD, as $QD_{h.pos}$, from the lateral center of the microlens using free-space collection apertures with NAs of \num{1} and \num{0.55} as well as lensed single mode fibers featuring an NA of \num{0.6}.}
    \label{fig:results_simulation}
\end{figure}

The observed statistical distribution of QD-microlens brightness can be understood by the integration of QD occupation probability with a horizontal displacement-dependent brightness function for the microlens shape. The occupation probability is modeled by the quantum dot area density and the occupation area with a Poisson distribution function. The displacement-dependent brightness function is ascertained through FDTD simulations, which describe the EE and ICE with respect to the horizontal displacement of QDs from the center of the microlens, as illustrated in FDTD simulations \cref{fig:results_simulation}e obtained for the RIE shape extracted from \cref{fig:luminescence}c. Given the pronounced correlation between ICE and LEE, it is assumed that predictions for ICE are comparable to the LEE. The detailed numerical modelling and analysis is described in supplemental \cref{sec:distribution_of_QDs_in_microlenses}.

According to this model, calculations for a sample comprising \num{800} microlenses, using the experimentally determined QD area density of \(\rho_{A} = \qty{1.6(2)}{}\times\qty{e11}{\text{QDs}\per\meter^2}\) and microlens radius of \qty{1270}{\nm} predict the following: \num{356} microlenses remain unoccupied by QDs, \num{288} microlenses contain a single QD, and \num{156} microlenses host two or more QDs. For a singular QD situated within the center of the microlens, i.e. $r_{C} \leq \qty{100}{\nm}$ (cf. \cref{fig:results_simulation}), the probability of experiencing exceptionally enhanced brightness is computed as \num{1} in \num{200} microlenses. This prediction is consistent with the observation of four bright microlenses out of \num{800} in the experimental data. For an assessment of the brightest microlenses the model gives good results. Nonetheless, the predicted number of unoccupied microlenses is only half of the experimentally observed figure. This may be attributed to the fact that QD located close to the microlens edge are quenched due to interference from surface states. Alternatively there may also deviations due to local fluctuations in QD area density \(\rho_{A}\) induced by variations in MBE growth conditions \cite{konishi2017spatial, babin2022full}. 

The determination of microlens occupation by multiple QDs solely through PL spectra presents considerable challenges. Nevertheless, the displacement-dependent efficiency function, as employed for this setup, specifically amplifies emissions from QDs located near the center of the microlens, implying that moderate occupation by multiple QDs is anticipated to exert negligible influence. For the purpose of bight entangled photon pair sources the impact of multiple QDs per lens is negligible as for this purpose resonant driving of specific QD resonances is used. This method suppresses the emission of undesirable QDs greatly. While the yield analysis can be modeled qualitatively, it is worth noting that the model is based on idealized conditions and therefore potentially ignores spatial inhomogeneities or variations in microlens shapes. A comparative analysis of the simulated brightness distribution for free-space optics and fiber-coupled collection methods featuring NAs of \num{0.55} and \num{0.60}, respectively, reveals that the experimentally observed QD brightness distribution, is modeled reasonably well by ICE but not by EE, cf. \cref{fig:results_simulation}e. A detailed discussion is found in supplemental \cref{sec:distribution_of_QDs_in_microlenses}. This underscores the necessity to consider the employed method of collecting the emitted photons of photonic micro devices, such as lensing into single mode fibers or free-space detection, as only then do the defined extraction characteristics and rates acquire practical significance.

Prospective improvements to the device yield analysis may include the implementation of high-resolution spatial mapping of QD distributions, as well as the advancement of simulations that integrate non-Poissonian effects. Such enhancements are anticipated to increase the precision of predictions and facilitate the optimization of light extraction efficiency in microlens-based photonic devices.

A comprehensive examination of the average fabricated monolithic microlens devices, as shown by the AFM profilometry in \cref{fig:luminescence}c, with respect to its performance in terms of EE, CE, and ICE, is conducted via FDTD simulations. The modeling is performed analogous to the idealized microlenses, cf. \cref{sec:concept}, but uses a 3D-shape derived by experimental profilometry instead of a hemispherical one. As such, the model retains two variables: $d_{QD-m}$ representing the distance between gold-back reflector and QD as well as the etch depth $Lens_{v.pos}$. The latter constitutes deviations of the lens height compared to \qty{1.35}{\um} by over- or under-etching compared using extended or shorted RIE exposure times, cf. \cref{sec:shape_transfer}. The isotropic etching effect, i.e. shrinking of the lateral dimensions (cf. \cref{sec:shape_transfer}), of the lens shape is however not considered. The simulation examines an $Lens_{v.pos}$ interval of \qty{-350}{\nm} to \qty{100}{\nm}. Analogous to the idealized lens shape, Fabry -P\'erot resonances in both parameters are observed. Due to the lower symmetry of the etched lens shape, more complex patterns of sub-resonances are apparent.\\

The investigation of the far-field emission pattern and the angular dependence of the EE has disclosed a diminished prominence of vertical light extraction compared to the hemispherical microlens configurations. Approximately only \qty{25}{\percent} of the EE is confined within collection angles corresponding to a NA of \num{0.25}, increasing to \qty{47}{\percent} at an NA of \num{1} for $Lens_{v.pos} = \qty{-212}{\nm}$  and $d_{QD-m} = \qty{127}{\nm}$. Consequently, the overall EE of the RIE Lens is approximately \qty{24}{\percent} lower compared to a hemispherical lens structure, which is attributed to its deviation from the optimal geometric configuration.
The fiber mode overlap CE, which is indicative of the mode matching with a target optical mode of a single-mode fiber, exhibits a periodic pattern analogous to that of the EE. Notably, the maximal CE value of \qty{84}{\percent} is actually higher than the value of \qty{72}{\percent} observed for the hemispherically shaped microlens. This can be attributed to a focussing effect of the experimental shape, due to higher refraction angles at its side. The ICE achieved its maximal value of \qty{21}{\percent} at well-defined Fabry-P\'erot oscillations $Lens_{v.pos}$ and $d_{QD-m}$. A broadened maxima of ICE of \qty{16}{\percent} is observed at a reduced height of the order of \qty{-200}{\nm}, implying that an extended etching process could further enhance the ICE by adjusting the profile toward regions of optimal compatibility between EE and CE.

In addition to the vertical positioning of the QD and the vertical lens position, the horizontal positioning of the QD is pivotal for optimal device performance. \cref{fig:results_simulation}e depicts the simulated EE as a function of lateral displacement, $QD_{\text{hor-pos}}$, ranging from \qty{0}{\nm} to the lens radius, for $Lens_{v.pos} = \qty{-212}{\nm}$ and $d_{QD-m}=\qty{127}{\nm}$. The findings show a periodic decline in EE from the center of the lens toward its periphery, a pattern consistently observed across all NA. This trend underscores the substantial influence of horizontal modal confinement on the coupling efficiency of the QD. Contrarily, the behavior for coupling to single-mode fibers ICE only exhibits high values for small lateral displacements of $QD_{\text{hor-pos}}$ from the microlens center. This effect can be attributed to the strong dependence of the CE on symmetric emission profiles, which are only obtained for QD near the lens center. This effect therefore also strongly suppresses multi-QD emission from the same microlens.

As discussed in detail in the supplemental material \cref{sec:distribution_of_QDs_in_microlenses}, this effect would allow for much higher QD densities of up to $\rho_{A} = \qty{3.2e13}{\text{QDs}\per\square\meter}$ to be employed in order to increase the yield of monolithic microlenses with randomly positioned QDs to values of up to \qty{30}{\percent}. However, the practical limits of local QD area density remain an open question. Previous studies suggest an upper bound for liquid droplet etched $\GaAs$ QDs of approximately $\rho_{A} = \qty{7.0e11}{\text{QDs}\per\square\meter}$ \cite{kruck2024critical}, for QDs grown in the Stranski-Krastanov mode this could however be higher. Further investigations are necessary to evaluate challenges such as QD clustering, size uniformity, growth kinetics, and potential degradation of optical properties. Addressing these challenges could pave the way for significant improvements in the yield of bright-photon sources in microlens-based photonic devices. Another approach to reach this goal would be to implement deterministic lateral positioning of QDs and microlenses \cite{gschrey2015highly}.

\begin{figure}[H]
    \centering
    \includegraphics[width=\textwidth]{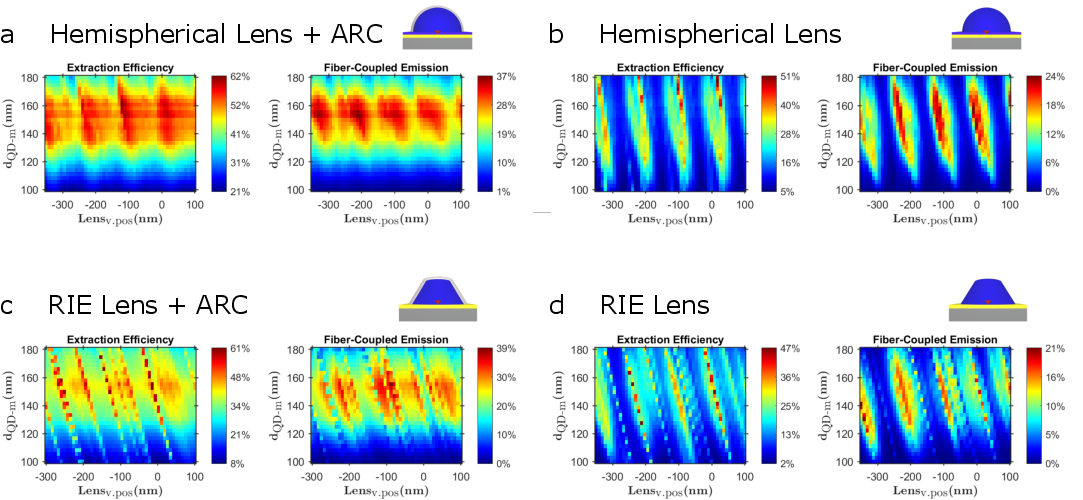}
    \caption{Finite-difference time-domain (FDTD) simulated extraction efficiencies of free-space and fiber-coupled emission of $\AlGaAs$ monolithic microlenses with hemispherical (a,b) and fabricated reactive ion etched (RIE)-lense  (c,d) shape. Additionally, shapes with and without anti-reflection coating (ARC) (a,c) and (b,d), respectively, are shown. Device modeling and simulations are performed as outlined in \cref{sec:conclusion}.}
    \label{fig:results_simulation_full}
\end{figure}

The role of an anti-reflection coating (ARC) and the deviations from the ideal shape, as elaborated above, become increasingly evident through a direct comparison of the EE and ICE of a hemispherical lens, both with and without ARC, and of the RIE lens, both with and without ARC.

The hemispherical lens equipped with ARC demonstrates overlapping periodic maxima in both EE and ICE, characterized by a width of $\leq\qty{50}{\nm}$. In contrast, in the absence of ARC, the periodic maxima become more pronounced, with narrower widths $\leq\qty{25}{\nm}$ and an expected general reduction in EE and ICE. The EE of the RIE Lens with ARC attains maxima similar to those of the hemispherical lens with ARC, albeit with significantly sharper widths, necessitating highly precise positioning of QDs within the structures. This requirement presents significant challenges to the fabrication process. As discussed above, the RIE lenses feature additional sub-resonances due to their reduced symmetry as compared to the idealized hemispherical ones. The application of an ARC coating can alleviate the pronounced resonant characteristics, resulting in comparable ICE-performance of both RIE and hemispherical microlenes.

The behavior of the RIE Lens without ARC closely mirrors that of the hemispherical lens without ARC, though with a less uniform periodic maxima distribution. In particular, the RIE Lens exhibits a specific vertical position, $Lens_{\text{vert-pos}} = \qty{-212}{\nm}$, where the high EE range is significantly broader on the order of \qty{40}{\nm}, a manufacturing precision achievable with common fabrication techniques. Optimization of the RIE Lens geometry, with precise control over $Lens_{\text{vert-pos}}$, $d_{QD-m}$, and $QD_{\text{hor-pos}}$, could approach the optical performance of hemispherical lenses, both with and without ARC, thereby enhancing light EE and positioning the RIE Lens as a viable alternative for the idealized hemispherical microlenses.

These simulation results show that coating monolithic microlenes with an ARC is beneficial to manufacturing precision requirements as well as to enhance the overall performance of the QD-microlens devices. Applying and verify these findings to practical devices is therefore the next logical step towards realizing compact, efficient and scalable entangled photon pair sources based on $\GaAs$ QDs.\\

%Experimental characterization of these devices reveal a QD spectroscopic enhancement of at least a factor of \num{200}.

\section{\label{sec:conclusion} Conclusion}
The demonstrated $\AlGaAs$ monolithic microlenses with embedded $\GaAs$ QDs can substantially enhance the extracted photon flux by a factor of at least \num{200} and therefore provide the basis for bright broadband sources of single and entangled photon devices. Through a combination of simulation and experimental optimization, monolithic microlenses with a shape close to an ideal hemisphere with a diameter of \qty{2.7}{\um} and height of \qty{1.35}{\um} are fabricated. The achieved light enhancement is comparable to other state-of-the-art designs of bright and broadband QD-photon sources \cite{Wang2019, Liu2019,nie2021experimental} and has recently successfully been employed for realization of an ultra-compact, fiber-coupled entangled photon pair source \cite{Langer2025}. The experimental findings are corroborated by simulations using AFM-determined 3D-profiles of the manufactured microlenses, predicting a maximal fiber-coupled extraction efficiency of about \qty{21}{\percent} when paired with lensed single mode fibers with a numerical aperture of \num{0.6}. Further simulations show that this value could be enhanced to values of up to \qty{39}{\percent} if anti-reflection coatings - such as $Al_2O_3$ - are employed.

A combination of temperature induced reflow of photo resist templates and carefully tuned shape transfer employing RIE is used to fabricate monolithic microlenses on a gold-coated substrate closely resembling hemispherical shapes. The efficient parallel processing of large arrays of microlenses using optical lithography and low-strain membrane transfer enables the scalable fabrication of these devices. Through these means, lenses with diameters between \qty{1.7}{\um} and \qty{5.0}{\um} are obtainable. The achievable aspect ratio can be tuned through a combination of resist thickness, photoresist template diameter and etching RF and ICP powers. In order to obtain the demonstrated high-quality QD-microlens devices resembling hemispherical shapes, the following set of parameters is employed: \qty{440}{\nm} resist thickness, \qty{140}{\celsius} reflow temperature for \qty{60}{\minute}, \qty{6}{\minute} etching time, \qty{155}{\nm} QD to gold layer thickness, \qty{110}{\watt} RF-Power, \qty{450}{\watt} ICP-Power and \qty{10}{\nm} $\mathrm{Al_{2}O_{3}}$ anti-oxidation coating.

The yield of the manufactured QD-microlens devices is estimated to a value of \num{1} in \num{200}. These findings are corroborated by statistical modelling combined with simulation results. The yield could potentially be enhanced dramatically by utilization of deterministic positioning of microlenses on top of single QDs \cite{gschrey2015highly} or by increasing the QD density. By employing Van-der-Vaals bonded QD-nanomembranes on a gold-coated $\GaAs$ substrate wafer, stable, $mm^2$-scale manufacturing of dense arrays (\qty{40000}{\per \square \mm}) of high homogeneity monolithic microlenses is achieved. \clearpage

\section*{\label{acknowledgment} Acknowledgment}
We thank the clean room team, especially Ronny Engelhard, of the Leibniz IFW Dresden for his efforts and expertise in clean room processing of samples and preparation of scanning electron beam and focused ion beam images. This work was funded by the German Federal Ministry of Education and Research (BMBF) projects QR.X, QUIET and QD-CamNetz (contracts no. 16KISQ016, 16KISQ094, and 16KISQ078).

\section*{\label{das} Data Availability Statement}
All data that support the findings of this study are included within the article (and any supplementary files).

\section{\label{sec:Suppl}Supplemental Information}

\subsection{Methods}
\label{sec:methods}

\subsubsection{Heterostructure Growth}
\label{sec:growth}

The $\AlGaAs$ heterostructure samples are grown using molecular beam epitaxy (MBE) on a $\GaAs$(100):Si substrate. The epitaxy process is initialized by a \qty{100}{\nm} $\GaAs$ buffer layer, which is then followed by a $\GaAs$/$\AlAs$ superlattice of \num{20} periods of \qty{3}{\nm} each, and a \qty{100}{\nm} $\GaAs$ buffer layer for surface smoothing. A \qty{30}{\nm} $\AlAs$ sacrificial layer is added for selective etching in the lift-off process. The $\AlGaAs$ membrane features a thickness of \qty{1570}{\nm}. The liquid droplet etched (LDE) $\GaAs$ QDs are embedded into this layer at a thickness of \qty{155}{\nm}. The LDE holes exhibit widths of \qty{40}{\nm} and depths \qty{12}{\nm}, and are filled by ca. \qty{8}{\nm} $\GaAs$. Details of the LDE $\GaAs$ can be found in Ref. \cite{keil2017solid}. QD density at the substrate center is $n_{QD} = \qty{16e10}{\text{QDs}\per\square\meter}$, as determined from atomic force microscopy (AFM) of an identical sample, that was stopped after LDE, cf. \cref{fig:dot_in_center}a.

\subsubsection{Membrane Transfer}

After growth, samples are cleaved into $4 \times 4 \, \text{mm}^2$ pieces and processed into membranes by selective wet etching of $\AlAs$ using hydrofluoric (HF) acid. Using the epitaxial lift-off (ELO) process pioneered by Yablonovitch \textit{et al.} \cite{yablonovitch1987extreme}, large substrates are  under-etched using a top-deposited tensile stressor to open and stabilize etch channels. This approach prevents clocking and therefore enables reliable sacrificial layer removal over large areas. The process was adapted by replacing Black-Wax with AZ1500 photoresist, which is stable against HF and available in standard clean-room environments. The photoresist is spin-coated to about \qty{1.5}{\um}. After etching in \qty{25}{\%} HF solution, membranes are transferred onto $6 \times 6 \, \text{mm}^2$ $\GaAs$(100) substrates with a \qty{70}{\nm} gold coating. Photoresist is removed using acetone, and heat-treated at \qty{130}{\degree C} for \qty{24}{h} to strengthen the Van der Waals bonds for microlens fabrication. This membrane transfer method avoids straining the material, as compared to methods such as gold-to-gold bonding, and therefore avoids strain induced shifting of the QD energy level structure. \\

\subsubsection{Monolithic Microlens Fabrication}
\label{sec:suppl_fabrication}

Cylindrical resist structures with diameters \qty{2}{}-\qty{5}{\um} are patterned on AZ1500 photoresist (thickness \qty{450}{}-\qty{1650}{\nm}) via maskless laser lithography. Exposure and development are optimized for each thickness seperately. Structures are thermally reflowed into lenslets  at \qty{130}{\degree C} for \qty{60}{min}, then etched into the semiconductor with argon and chlorine gases, cf. \cref{sec:fabrication} for details. A polymethyl methacrylate (PMMA) bond layer is used to assure cooling of the substrate inside the reactive ion etching machine \cite{giehl2003deep}. Etch rates for the resist and $\AlGaAs$ membranes are calculated by profilometry, see below, before and after etching, and after resist removal. Etch time for complete resist removal and overetching are calculated from these investigations. All microlens samples receive a \qty{10}{\nm} layer of $\mathrm{Al_{2}O_{3}}$ to prevent oxidation and material degradation.

\begin{figure}[H]
    \centering
    \includegraphics[width=\textwidth]{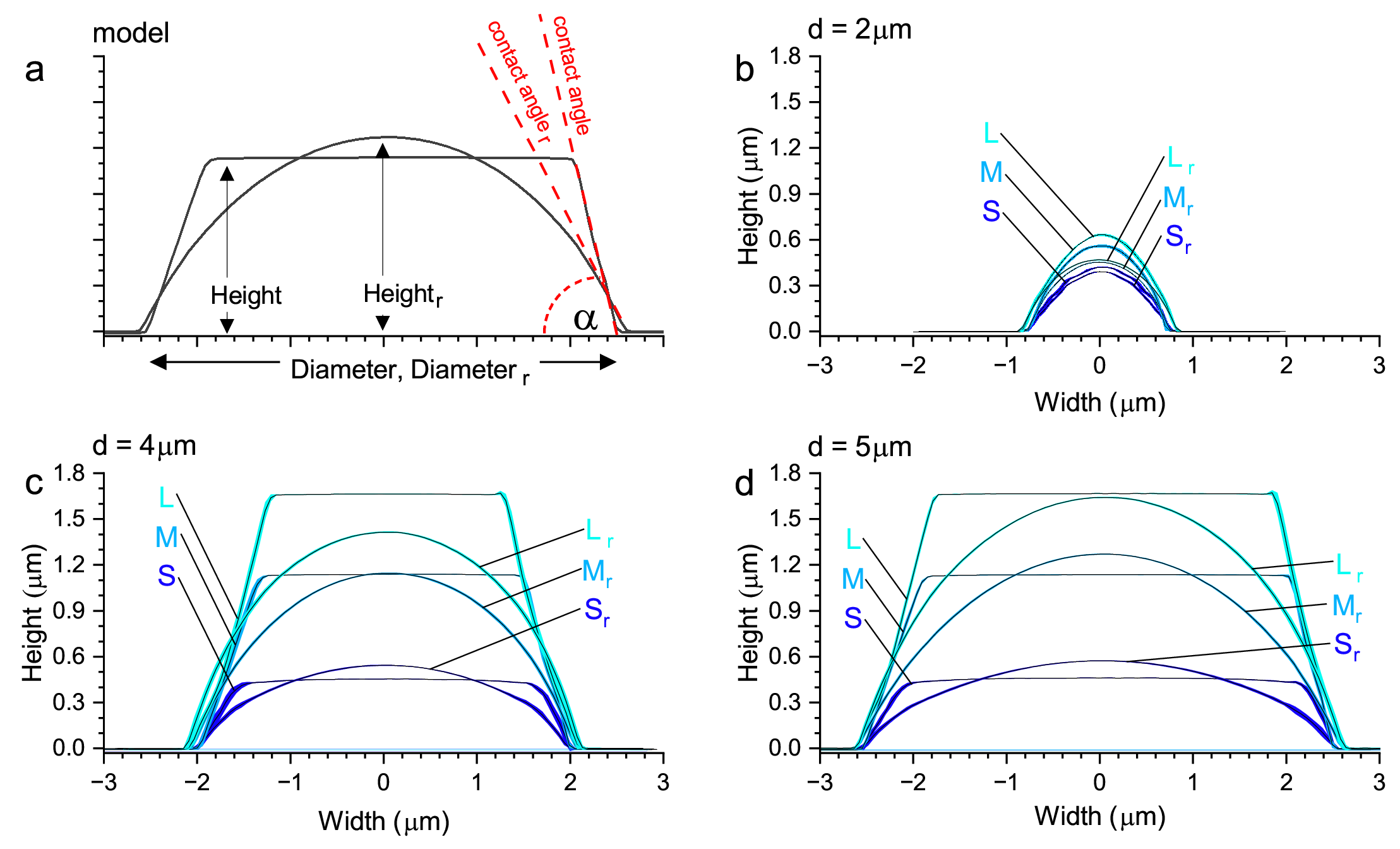}
    \caption{a) Definition of the contact angle, thickness, height, and diameter of structures before and after (r) thermal reflow. 
    (b-d) Average height profiles determined by atomic force microscopy (AFM) of \num{10} fabricated structures with nominal photoresist template diameters $d$ of \qty{2}{\um}, \qty{4}{\um}, and \qty{5}{\um}, respectively. 
    The structures were fabricated using spin-coated resist layers with thicknesses of \qty{1.66}{\um} ($L$), \qty{1.14}{\um} ($M$), and \qty{0.45}{\um} ($S$). 
    Thickness measurements were conducted using AFM both before and after (r) employing thermal reflow.}
    \label{fig:resistfabrication_all}
\end{figure}

The determined height profiles of photoresist microlens templates before and after reflow are summarized in \cref{fig:resistfabrication_all}  as a function of nominal template diameters and employed resist thicknesses. The corresponding resist thicknesses are summarized in \cref{tab:resist_dimensions1}. As illustrated in \cref{fig:resistfabrication_all}a, the contact angle of the resist is defined as the angle at which the resist structure meets the semiconductor substrate. The obtained contact angles are summarized in \cref{tab:resist_dimensions2}. While before the reflow process this angle is determined by the photolithography process, i.e. by the proximity effect, after reflow it is largely independent from the template diameter. This can be understood by considering that this angle is a product of the surface energy of the liquid resist during the reflow process. This effect limits the achievable shape of the resist lenslets and therefore also how close the monolithic microlenes shape can be made to resemble that of a hemispherical one after transfer into the $\AlGaAs$ material. \\

\begin{table}[ht]
    \centering
    \caption{Resist Thicknesses}
    \begin{tabular}{@{}lcccccc@{}}
        \toprule
        Diameter \textit{d} & $S$ & $S_{r}$ & $M$ & $M_{r}$ & $L$ & $L_{r}$ \\ \midrule
        \qty{2}{\um} & \qty{393(2)}{\nm} & \qty{420(3)}{\nm} & \qty{453(4)}{\nm} & \qty{564(8)}{\nm} & \qty{471(4)}{\nm} & \qty{634(8)}{\nm}       \\
        \qty{3}{\um} & \qty{493(1)}{\nm} & \qty{449(1)}{\nm} & \qty{958(3)}{\nm} & \qty{1138(1)}{\nm} & \qty{1173(7)}{\nm} & \qty{1658(2)}{\nm}       \\
        \qty{4}{\um} & \qty{546(2)}{\nm} & \qty{462(1)}{\nm} & \qty{1148(2)}{\nm} & \qty{1143(2)}{\nm} & \qty{1424(5)}{\nm} & \qty{1662(4)}{\nm}       \\
        \qty{5}{\um} & \qty{574(3)}{\nm} & \qty{465(3)}{\nm} & \qty{1274(2)}{\nm} & \qty{1143(2)}{\nm} & \qty{1646(7)}{\nm} & \qty{1665(2)}{\nm}       \\ \bottomrule
    \end{tabular}
    \label{tab:resist_dimensions1}
\end{table}

\begin{table}[ht]
    \centering
    \caption{Contact Angles}
    \begin{tabular}{@{}lcccccc@{}}
        \toprule
        Diameter \textit{d} & $S$ & $S_{r}$ & $M$ & $M_{r}$ & $L$ & $L_{r}$ \\ \midrule
        \qty{2}{\um} & \qty{42}{\degree} & \qty{42}{\degree} & \qty{51}{\degree} & \qty{50}{\degree} & \qty{54}{\degree} & \qty{50}{\degree}       \\
        \qty{3}{\um} & \qty{46}{\degree} & \qty{40}{\degree} & \qty{56}{\degree} & \qty{55}{\degree} & \qty{54}{\degree} & \qty{57}{\degree}       \\
        \qty{4}{\um} & \qty{44}{\degree} & \qty{33}{\degree} & \qty{60}{\degree} & \qty{51}{\degree} & \qty{53}{\degree} & \qty{56}{\degree}       \\
        \qty{5}{\um} & \qty{45}{\degree} & \qty{31}{\degree} & \qty{61}{\degree} & \qty{47}{\degree} & \qty{55}{\degree} & \qty{58}{\degree}       \\ \bottomrule
    \end{tabular}
    \label{tab:resist_dimensions2}
\end{table}

\subsubsection{Profilometry}

Shape profiles of resist and monolithic microlenses are characterized by atomic force microscopy (AFM) featuring a spatial resolution of \qty{80}{\nm}. Height profiles of the photoresist are determined by sampling  
\num{10} structures, while monolithic microlenses involve \num{15}. Profiles are overlaid and averaged to assess height and width variations. Time-resolved etch profiles correspond to distinct runs using a consistent resist template. Etch selectivity is determined by a stylus profiler performing 10 measurements of etched bars before and after etching, and after photoresist removal, with averaged results determining selectivity and etch rates.\\

\subsubsection{Spectroscopy}

The photoluminescence (PL) emission of $\GaAs$ QDs is measured in a low-temperature He-flow cryostat at \qty{5}{\kelvin}. QDs are excited to saturation of the QD neutral exciton ($X^0$), cf. Ref. \cite{Hopfmann2021a}, with a \qty{635}{\nm}  continuous-wave laser using a 90:10 beam splitter. Excitation and luminescence collection are conducted with an NIR objective with NA \qty{0.55}. Emitted luminescence is filtered by a \qty{700}{\nm} long-pass filter and resolved by a \qty{1200}{g/mm} diffraction grating and \qty{50}{\um} slit. Random QDs are selected from planar and membrane samples. 800 microlenses are scanned, averaging 5 measurements each. PL spectra are integrated over \qty{775}{}-\qty{785}{\nm} and the resulting values histogrammed to show emission intensity distribution, cf. \cref{fig:luminescence}f. The $X^0$ emission is modeled by a Voigt-profile using an automated computer program. The $X^0$ emission energy is extracted in order to investigate the statistical distribution of the QD emission characteristics.\\

\subsubsection{Finite-Difference Time-Domain Simulations}

Finite-difference time-domain (FDTD) simulations use commercial software. Models consist of a \qty{2}{\um} thick $\GaAs$ substate coated by a \qty{500}{\nm} gold layer. The hemispherical lens model includes a $\AlGaAs$ layer with thickness $d_{QD-m}$ between QD and the gold-mirror, as well as a hemisphere with diameter $D_{Lens}$. The QD is positioned in the center of the hemisphere, see also \cref{fig:concept}e. The anti-reflection coating is a second enclosing hemisphere with diameter $D_{ARC}=D_{Lens}+\qty{112}{\nm}$.
The modeling of the fabricated RIE microlens is based on a negative height profile, placed as an etch mask on a $\AlGaAs$ slab with \qty{3}{\um} thickness on gold, thereby forming the microlens. Both the hemispherical lens and negative mold can be shifted horizontally while keeping the QD position constant,to simulate varying dimensions of the lens. The microlens vertical displacement parameter $Lens_{v.pos}$ shifts the basis of the negative etch mold of the microlens profile away from the gold-layer. This effect of this parameter is that for positive values the membrane is not completely etched away outside the lens center, effectively leading to a membrane-lens structure, similar to the hemispherical lens model. Negative values result in over-etching, reducing the lens height and diameter. The QD is modeled as an in-plane dipole point source emitting at a wavelength of \qty{780}{\nm}.

\subsection{Yield Analysis Of QD-Microlenses}
%Distribution of QDs in microlenses
%
\label{sec:distribution_of_QDs_in_microlenses}

\begin{figure}[H]
    \centering
    \includegraphics[width=\textwidth]{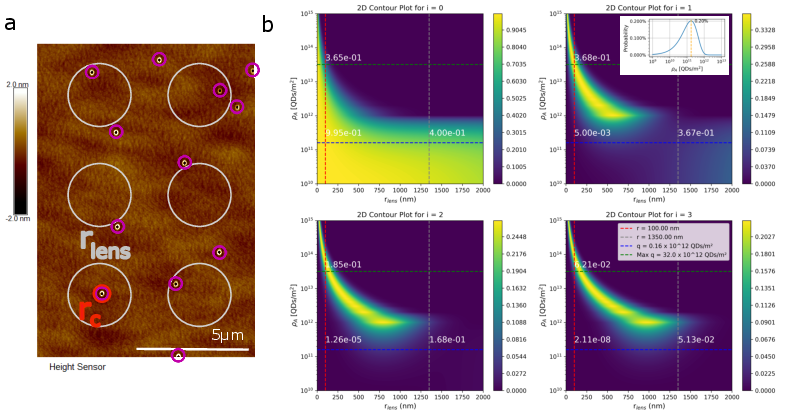}
    \caption{a) Atomic force micrograph (AFM) surface height profiles of an unprocessed $\AlGaAs$ sample with droplet etched holes at its surface (purple circles) illustrating microlens integration (gray circles) over randomly localized liquid droplet etched QDs. The microlens radii $r_\text{lens} =$ \qty{1270}{\nm}, spacing $d_\text{sp} =$ \qty{5}{\um}, and critical radii $r_C =$ \qty{100}{\nm} for high extraction efficiency, cf. \cref{fig:results_simulation}, are denoted. b) Simulation of QD occupation probability of number of QDs $i$ in a circular area with radius r as a function of QD area density $\rho_A$, using Poisson statistics. The blue line illustrates the $\rho_A$ of \qty{0.16e12}{\text{QDs}\per\square\meter} present in fabricated devices of \cref{sec:results}, while the green line represents the optimal value of \qty{3.2e13}{\text{QDs}\per\square\meter}. The gray and red dashed lines illustrate the lens and critical radii $r_\text{lens}$ and $r_C$ of the fabricated lenses, respectively. The occupation probability of a single QD within $r_C$ and no further QD within $r_m$ is shown as an inset.}
    \label{fig:dot_in_center}
\end{figure}

Due to the random positioning of QDs on the sample surface in the LDE process \cite{keil2017solid}, it can be assumed that the spatial QD position follows a Poisson distribution with uniform QD area density $\rho_{A} \;[\qty{}{\text{QDs}\per\square\meter}]$ across the microlens array's surface. The parameter $\rho_{A}$ is derived from AFM measurements of LDE holes on the sample surface, which are presumed to provide an analogous result $\rho_{A}$ to that of the buried QDs under identical growth conditions. The microlens array is characterized by a grid configuration with a spacing $d_{sp}$ of \qty{5}{\um}, comprising circular microlenses each with a radius $r_{lens}$. This is illustrated in \cref{fig:dot_in_center}a. The expected value of how many QDs are found per microlens is given by the product of $\rho_{A}$ and the lens surface area $A$ of given the radius $r$:

    \begin{equation} 
    N\;=\;\rho_{A}\cdot\pi\cdot r^{2}\;=\;\rho_{A}\cdot A
    \label{eq:N}
    \end{equation}

The probability of $i$ QDs in the area $A$ for $\rho_{A}$ is calculated by the Poisson distribution:
    \begin{equation} 
    P(i)\;=\;\frac{(N)^{i}e^{-N}}{i!}\;=\;\frac{1}{E(i)}
    \label{eq:P(i)}
    \end{equation}

The reciprocal of the probability, $E(i)$, indicates the average number of microlenses that must be examined to locate a single microlens containing $i$ QDs in $A$. According to \cref{eq:P(i)}, the likelihood of finding $i$ QDs within a microlens is unaffected by the configuration of the microlens array, i.e., its form and spacing $d_{sp}$. These design parameters solely influence the area density of lens yield with $i$ QDs, without altering the probabilities per lens. The computed probabilities for parameters $i\:=\:0;1;2;3$, $r=\qty{0}{}\!-\!\qty{2000}{\nm}$, and $\rho_{A}=\qty{e10}{}\!-\!\qty{e15}{\text{QDs}\per\square\meter}$ are presented in \cref{fig:dot_in_center}b.\\

The red dotted line delineates the radius $r_{c}$ with \qty{100}{\nm}, signifying the radius for a high EE as determined by the displacement-EE function referenced in \cref{fig:luminescence}f. The gray dotted line indicates $r_{lens}$ with \qty{1350}{\nm}, which delineates the area pertaining to the fabricated microlenses of \cref{sec:results}. The parameter $\rho_{A}$, for the examined sample, is indicated by \qty{0.16e12}{\text{QDs}\per\square\meter}. Its corresponding value is denoted by the blue horizontal line, and the points of intersection yield the probabilities of $i$ QDs within $r_m$. The probability of precisely one QD ($i=1$) being present within the radius of high EE is $P(1, r_{c})=\qty{0.5}{\percent}$, culminating in $E(1, r_{c})=200$. For a sample size of 800, this results in an expectation to find 4 microlenses exhibiting high EE. The probabilities associated with the presence of multiple QDs within this radius, $P(i>1, r_{c})<\qty{0.002}{\percent}$, are insignificantly small.
The single QD purity of the microlens array is computed as described by $P(1, r_{lens})=\qty{36.0}{\percent}$. The additional quantitative values are denoted by $P(0, r_{lens})=\qty{44.5}{\percent}$, $P(2, r_{lens})=\qty{14.6}{\percent}$, $P(3, r_{lens})=\qty{3.9}{\percent}$. Consequently, for a sample size of 800, there are approximately 356 microlenses with no QD, 288 with a single QD, 117 with two QDs, and 39 with three or more QDs. In conclusion, the likelihood of bright microlenses is solely influenced by the QD area density, as determined by the MBE LDE $\GaAs$ quantum dot growth process, and by the radius of high EE, which is governed by the characteristic shape of the microlens.

Assuming a fixed microlens geometry with inadequate suppression of multi-QD emission, the yield can solely be optimized through the QD area density. The inset of Figure \cref{fig:dot_in_center}b illustrates the likelihood $P(1, r_{C})*P(0, r_{m})$ of attaining a single QD with high brightness within $r_C$, with no other QD present within $r_{m}$. The findings, specific to each microlens configuration, suggest an optimal \(\rho_{A}\) of \(\qty{1.75e12}{\text{QDs}\per\square\meter}\) under these premises.

In scenarios where the shape of the microlens is fixed and the suppression of multi-QD emission is adequately achieved, i.e. by spectroscopic means, the findings in \cref{fig:dot_in_center}b suggests that the yield of luminous microlenses could significantly increase by higher $\rho_{A}$. The horizontal green line denotes the value of $\rho_{A} = \qty{3.2e13}{\text{QDs}\per\square\meter}$, which corresponds to the maximum probability of \(P(1, r_{c}) = \qty{36.8}{\percent}\). The comparatively minor influence of $P(0, r_{lens})$ can be disregarded in this context. Consequently, the device-yield of QD-microlenses could be notably enhanced, with one out of three microlenses exhibiting high EE. At elevated values of $\rho_{A}$, the likelihood of single-QD occupation diminishes due to the increased probability of multi-QD occupation, as noted in $P(i = 1; 2; 3; \dots, r_{c})$, cf. \cref{fig:dot_in_center}b.

\begin{figure}[H]
    \centering
    \includegraphics[width=\textwidth]{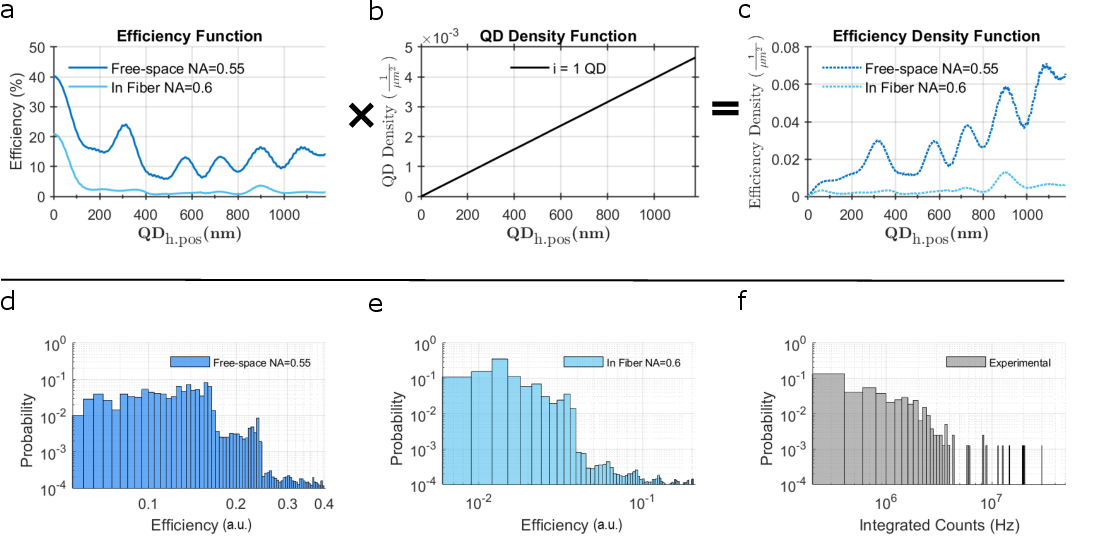}
    \caption{
    a) Simulated extraction efficiencies into free-space collection optics featuring a numerical aperture (NA) of \num{0.55} and into lensed 780HP single-mode fibers (NA of \num{0.6}) as a function of the horizontally displaced $QD_{h.pos}$ of the QD from the center of the etched microlens. 
    b) QD density function as a function of $QD_{h.pos}$ given the QD area density $\rho_{A} = \qty{16e10}{\text{QDs}\per\square\meter}$. 
    c) QD efficiency density function versus $QD_{h.pos}$, derived by multiplication of efficiency and QD density functions. 
    d) Statistical representation of Monte-Carlo simulation results for free-space collection (EE) using a NA of \num{0.55} as a function of the extraction efficiency. 
    e) Statistical representation of Monte-Carlo simulation results for lensed collection into a single mode fiber (ICE) using a NA of \num{0.6} as a function of the extraction efficiency. Details of the simulation can be found in the text.
    f) Experimentally determined brightness distribution of 800 microlenses, cf. \cref{sec:results}.}
    \label{fig:result_simulation_comp}
\end{figure}

The statistical distribution of microlens LEE, as illustrated in \cref{fig:luminescence}f, can be numerically investigated through Monte Carlo simulations. This methodology utilizes the efficiency function for the lateral QD position, obtained from FDTD simulations, alongside the spatial distribution function that characterizes the arrangement of QDs within a circular area of radius $r$, as previously discussed, cf. \cref{fig:result_simulation_comp}a,b. By multiplicative combining these functions, the efficiency density function shown in \cref{fig:result_simulation_comp}c is obtained. By drawing large sets of \num{1e8} random efficiency values distributed according to this efficiency density distribution, the Monte Carlo simulation results are obtained. Details on the Monte Carlo methodology can be found in Ref. \cite{thomopoulos2012essentials}. The obtained numerical efficiency distributions are depicted in  \cref{fig:result_simulation_comp}d-e for both free-space (EE) and lensed fiber collection (ICE) methods featuring NAs of \num{0.55} and \num{0.60}, respectively. The simulation results are compared to the experimental ones as shown in \cref{fig:result_simulation_comp}d-f. While the experimental count rates and theoretical efficiencies are clearly correlated, the experimental setup efficiency characteristics are not explicitly known. Therefore, the x-axis scaling factor between the experimental and theoretical curves is an undetermined and any comparison between these results is limited to a qualitative analysis. A comparison between the curves reveals, that the simulated collection through a lensed fiber closely represents the experimental curve, while the free-space collection shows significant deviation. At first glance this may be surprising, since the experiments are performed using a free-space spectroscopy setup, cf. \cref{sec:methods}. One has to, however, consider that the simulation of the free-space collection does not take into account how well the collected intensity can be collimated. Due to the fact that the experimental setup relies on collimation and refocusing the light into a spectrometer some distance (ca. \qty{1.5}{\metre}) away, only light which is collimated within narrow dispersion angles can be detected. This is conceptually very similar to the effect of matching the mode field of a single mode fiber. Using this relation, the experimental results can be fully qualitatively explained by the presented modeling. As such, it is found that up to about a count rate of \qty{3}{\MHz} (\qty{3}{\percent} efficiency) the experimental (ICE simulated) probability distribution is quite linear. Beyond that point, the probability drops by about two orders of magnitude. By associating this characteristic drop in both curves, it can be roughly estimated that for the best QD-microlens devices an extraction efficiency into a collimated mode of about \qty{30}{\percent} is achieved. This value is in good agreement with the simulated expectations, cf. \cref{sec:results}, where ICE efficiency values of up to \qty{39}{\percent} are obtained in the ideal case.\\

In summary, the presented yield estimation of microlenses with statistically positioned QDs shows that there is significant optimization potential. Specifically, by employing higher QD densities of values of about \qty{32e12}{\text{QDs}\per\square\meter} would increase the current yield of \num{1} in \num{200} to values of \num{1} in \num{3}. Furthermore, by employing statistical analysis by combined FDTD and Monte Carlo simulations, the expected statistical yield distribution and efficiencies of monolithic QD-microlens devices can be estimated qualitatively. Nevertheless, accurate spatial positioning of QDs within the lateral center of the monolithic microlenes could pave the way for yields close to \num{1}, although with increased fabrication complexity.\\

\printbibliography

\end{document}